\documentclass[twocolumn,preprintnumbers,amsmath,amssymb]{revtex4-1}
\usepackage{amsmath, amssymb}

\usepackage{graphicx}
\usepackage[version=3]{mhchem} 
\usepackage{algorithm}
\usepackage{algorithmic}

\newcommand{\commentoutA}[1]{}

\begin{document}
    
    \preprint{LA-UR-23-22052}
    
    \author{Yu Zhang}
    \email{zhy@lanl.gov}
    \author{Marc J. Cawkwell, Christian F.A. Negre}
    \affiliation{Theoretical Division, Los Alamos National Laboratory, Los Alamos, New Mexico 87545}
    
    \author{Oscar Gr{\aa}n\"{a}s}
    \affiliation{Department of Physics and Astronomy, Uppsala University, Box 516, SE-751 20 Uppsala, Sweden}
    
    \author{Anders M. N. Niklasson}
    \email{amn@lanl.gov}
    \affiliation{Theoretical Division T-1, Los Alamos National Laboratory, Los Alamos, New Mexico 87545, USA}
    
    \date{\today}
    
    \title{Extended Lagrangian Born-Oppenheimer Molecular Dynamics with DFT+U}

    \begin{abstract}
        Extended Lagrangian Born-Oppenheimer molecular dynamics (XL-BOMD) [Phys.\ Rev.\ Lett.\ vol.\ {100}, 123004 (2008)] is combined with Kohn-Sham density functional theory (DFT) using a DFT+U correction based on the Hubbard model. This combined XL-BOMD and DFT+U approach allows efficient Born-Oppenheimer molecular dynamics simulations with orbital-dependent corrections beyond regular Kohn-Sham density functional theory.  The extended Lagrangian formulation eliminates the need for the iterative self-consistent-field optimization of the electronic ground state prior to the force evaluations, which is required in regular direct Born-Oppenheimer molecular dynamics simulations. This method provides accurate and stable molecular trajectories, while reducing the computational cost per time step. The combined XL-BOMD and DFT+U approach is demonstrated with molecular dynamics simulations of a nitromethane molecular liquid and a system of solid nuclear fuel, UO$_2$, using self-consistent-charge density functional based tight-binding theory.
    \end{abstract}
    
    \maketitle
    
    \section{Introduction}
    
    Quantum-mechanical Born-Oppenheimer molecular dynamics (QMD) simulations based on Kohn-Sham density functional theory (KS-DFT) and the local density or generalized gradient approximation
    \cite{hohen,KohnSham65,RParr89,RMDreizler90,JPerdew92} are widely considered as a gold standard for molecular dynamics simulations 
    \cite{MKarplus73,CLeforestier78,RCar85,DRemler90,MCPayne92,TAArias92,RBarnett93,GKresse93,GGalli96,HBSchlegel01,MTuckerman02,PCarloni02,JMHerbert04,PPulay04,marx_hutter_2009,BKirchner12,MCawkwell12,MArita14}. 
    However, QMD simulations based on KS-DFT have limitations in capturing the behavior of systems with strong electron correlation. They may fail to accurately predict properties such as the existence of a band gap or the number of valence electrons, leading to incorrect characterization of a material's physical nature and its response properties.
    Furthermore, QMD simulations based on first principles KS-DFT have a high computational cost, as the fully relaxed electronic ground state must be determined prior to the force evaluation for each new atomic configuration. This involves constrained iterative charge optimization of the nonlinear Kohn-Sham energy functional. This process limits the accessible simulation time and size of systems that can be studied. The nonlinearities of the KS-DFT functional can also cause instabilities with non-conservative forces and a drift in the total energy. This limitation if of particular significance in QMD simulations using reduced complexity solvers that are needed to study large systems \cite{MCawkwell12,MArita14,TOtsuka16,ANiklasson16,THirakawa17,LDMPeters17,CNegre23} or for QMD simulation using specialized AI-hardware with low-precision floating-point operations \cite{JFinkelstein21B}. In all these cases the effect of numerical approximations can be magnified by the non-linearities of the Kohn-Sham functional and the associated iterative charge optimization.
    
    Strong electron correlation and a high computational cost are often interrelated problems. Materials with heavy elements and narrow bands pose computational challenges due to their large number of electrons per atom and difficulties in finding the relaxed electronic ground state solution. Also, these materials often require a theory level beyond regular KS-DFT to account for their strong electron correlation. Despite their close connection, these two problems have mainly been treated separately.
    
    One approach to address the problem of strong electron correlation is to incorporate DFT+U correction terms based on the Hubbard model \cite{VIAnisimov91,VIAnisimov97,AILichtenstein95,SLDudarev98,HKulik15}. This method approximates the effects of electron correlation through a semi-empirical and tunable correction term added to the Kohn-Sham energy functional. 
    
    To tackle the issue of high computational cost for QMD simulations, a framework for extended Lagrangian Born-Oppenheimer molecular dynamics (XL-BOMD) was recently introduced \cite{ANiklasson06,ANiklasson08,ANiklasson09,PSteneteg10,GZheng11,MCawkwell12,JHutter12,LLin14,MArita14,LLin14,PSouvatzis14,ANiklasson14,KNomura15,AAlbaugh15,CNegre16,ANiklasson17,JBjorgaard18,ANiklasson20}. This approach, inspired by Car-Parrinello molecular dynamics \cite{RCar85,DRemler90,GPastore91,FBornemann98,DMarx00,JHutter12}, includes extended electronic degrees of freedom alongside the nuclear degrees of freedom as dynamical variables. When combined with an approximate \textit{shadow} Born-Oppenheimer potential energy surface, XL-BOMD can avoid the computational overhead of the iterative electronic ground state optimization and the stability problems caused by non-conservative forces, providing physically accurate trajectories at only a fraction of the cost of regular direct Born-Oppenheimer molecular dynamics simulations \cite{ANiklasson21b}.
    
    The two methods, DFT+U and XL-BOMD, have so far only been used separately. The main purpose of this article is to present a framework for QMD simulations that combines DFT+U and XL-BOMD. In this way we can reduce the computational cost of QMD simulations also for some materials with electron correlation effects beyond the reach of regular KS-DFT. The construction of the combined framework for DFT+U and XL-BOMD presents an example of a fairly general approach that can be applied also to other corrections of the Kohn-Sham functional besides the DFT+U term, for example, self-interaction corrections (SIC) \cite{SIC,ULundin01,MPedersen17,JPerdew15}. The DFT+U term may also serve as a tunable correction that could be used in machine learning approaches to adjust, for example, the polarizability of molecular systems in atomistic simulations using approximate DFT or Hartree-Fock methods \cite{MDewar77,MDewar85,MElstner98,MFinnis98,JStewart13,CBannwarth18,PDral19,WMalone20,ZGuoqing20,CBannwarth20,BHourahine20}. This can be achieved with a much lower computational overhead with the combined DFT+U and XL-BOMD approach.
    
    First we present KS-DFT using a density-matrix formulation and the orbital-dependent DFT+U correction term. We then introduce extended Lagrangian Born-Oppenheimer molecular dynamics (XL-BOMD)
    based on an approximate shadow potential energy surface for the DFT+U corrected Kohn-Sham
    density-matrix functional. The integration of the extended electronic equations of motion is discussed 
    in terms of a Krylov subspace approximation \cite{ANiklasson20,ANiklasson20b}.
    We then demonstrate QMD simulations using the combined XL-BOMD and DFT+U approach for a molecular system of liquid nitromethane and a solid of nuclear fuel, (UO$_2$), using self-consistent charge density functional based tight-binding theory (SCC-DFTB)
    \cite{MElstner98,MFinnis98,BHourahine20} before we present a summary and a discussion at the end.

    \section{Kohn-Sham density-matrix functional theory}
    
    Density functional theory is a cornerstone of electronic structure theory \cite{hohen,KohnSham65,RParr89,RMDreizler90,JPerdew92}.  
    KS-DFT in combination with the local density or generalized gradient approximations of the exchange-correlation energy is a computationally efficient and widely used formulation of DFT.
    KS-DFT is normally formulated in terms of the electron density. However, if we assume that all operators and potentials in KS-DFT are represented in some finite (atomic-orbital-like) basis set, $\{\phi_i({\bf r})\}_{i = 1}^N$, then a formulation based on the effective single-particle density matrix and density matrix energy functions is a more natural choice compared to the electron density. This is in analogy to Hartree-Fock theory \cite{Roothaan,RMcWeeny60} and is particularly useful when we introduce the orbital-dependent DFT+U energy correction. 
    
    In a finite basis-set representation, with $N$-basis functions), the ground state electronic structure in spin-independent KS-DFT can be described by the single-particle density matrix, 
    $\varrho_0 \in {\mathbb R}^{N \times N}$, that
    is given from a constrained density-matrix minimization ($\varrho \in {\rm C}$) of a matrix function,
    \begin{equation}\label{Dmin}
        {\displaystyle \varrho_{\rm 0} = \arg \min_{\varrho \in {\rm C}}\left\{ F_{\rm KS}[\varrho] + 2{\rm tr}[v_{\rm ext} \varrho] \right\}}.
    \end{equation}
    Here we assume that $F_{\rm KS}[\varrho]$ is the matrix-function approximation of the Kohn-Sham ensemble respresentation of the universal functional \cite{NMermin65,RParr89}
    in DFT at some chosen electronic temperature, $T_e \ge 0$. The density matrix constraints, $\rho \in C$, will be described below in Eq.\ (\ref{Cnstr}).
    The matrix elements of the external potential, $v_{\rm ext} \equiv v_{\rm ext}({\bf R}) \in {\mathbb R}^{N \times N}$, for the ions at positions ${\bf R} = \{{\bf R}_I\}$, are given by
    \begin{equation}
        {\displaystyle \left\{v_{\rm ext}({\bf R})\right\}_{ij} = \int \phi_i^* ({\bf r}) v_{\rm ext}({\bf R},{\bf r}) \phi_j({\bf r}) d{\bf r}}.
    \end{equation}
    The ensemble Kohn-Sham energy function, $F_{\rm KS}[\varrho]$, which is given at some electronic temperatures, $T_e \ge 0$, can be written as
    \begin{equation}\begin{array}{l}
            {\displaystyle F_{\rm KS}[\varrho]  =   2{\rm tr}[t_{\rm s}\varrho] + 2 \sum_{ij,kl} \varrho_{ij}\gamma_{ij,kl} \varrho_{kl} + E_{\rm xc}\left[\rho \right] }
            {\displaystyle + E_{\rm ent}(f)}, \\
        \end{array}
    \end{equation}
    where 
    \begin{equation}\label{K}
        {\displaystyle \{t_{\rm s}\}_{ij} = \int \phi_j^*({\bf r}) \left(-\frac{1}{2} \nabla^2\right)  \phi_i({\bf r})d{\bf r}},
    \end{equation}
    are the single-particle kinetic energy matrix elements, and
    \begin{equation}\label{Gamma}
        {\displaystyle \gamma_{ij,kl} = \iint \frac{\phi_i^*({\bf r})\phi_j({\bf r}) \phi_k^*({\bf r'})\phi_l({\bf r'})}{\vert {\bf r-r'}\vert } d{\bf r} d{\bf r'}},
    \end{equation}
    are the two-electron (Coulomb) integrals, and
    \begin{equation}
        {\displaystyle E_{\rm ent}(f) = 2 k_B T_e \sum_i \left( f_i \ln f_i + (1-f_i)\ln(1-f_i)  \right)} ,
    \end{equation}
    is the single-particle entropy contribution to the free energy, where we assume a double fractional occupancy of all the states.
    Here the occupation numbers $f_i \in [0,1]$ and $E_{\rm xc}\left[\rho \right]$ is the exchange-correlation energy functional that is approximated
    using the local density or the generalized gradient approximation. 
    $E_{\rm xc}\left[\rho \right]$ depends on the density $\rho({\bf r})$, which here is determined by the density matrix and the basis set, i.e.,
    \begin{equation}
        {\displaystyle \rho({\bf r}) = 2\sum_{i,j}^N \varrho_{i,j} \phi_i^*({\bf r}) \phi_j({\bf r})}.
    \end{equation}
    Because of this direct dependency on the density matrix we can alternatively use the notation $E_{xc}[\varrho] \equiv E_{\rm xc}[\rho]$.
    The two-electron integrals $\gamma_{ij,kl}$ are never calculated explicitly. Instead we use a contraction corresponding to the Hartree potential, $v_{\rm H}[\varrho]$, with matrix elements, 
    \begin{equation}\label{VCoulomb}
        {\displaystyle \{v_{\rm H}[\varrho]\}_{ij} =  \sum_{kl} \left(\gamma_{ij,kl} \varrho_{kl} + \varrho_{kl}\gamma_{kl,ij}\right)},\\
    \end{equation}
    which can be calculated, for example, with an Ewald summation for periodic boundary conditions.
    The Kohn-Sham free-energy matrix function can then be expressed as
    \begin{equation}\label{KSFunc}
        {\displaystyle F_{\rm KS}[\varrho]  =   2{\rm tr}[t_{\rm s} \varrho] + {\rm tr}[\varrho v_{\rm H}[\varrho]]  + E_{\rm xc}\left[\varrho \right] }
        {\displaystyle + E_{\rm ent}(f)}. \\
    \end{equation}
    The density-matrix minimization in Eq.\ (\ref{Dmin}) is performed under the costraints ($\varrho \in {\rm C}$), which require that
    \begin{equation}\label{Cnstr}
        \begin{array}{l}
            {\displaystyle  \varrho = \sum_i f_i c_i c_i^\dagger, ~~}
            {\displaystyle 2\sum_i f_i = N_{e}, ~~ f_i \in [0,1], }\\
            {\displaystyle \sum_{i,j} c_i^\dagger s_{i,j} c_j = \delta_{i,j}, ~~ }
            {\displaystyle s_{ij} = \int \phi_i^*({\bf r}) \phi_j({\bf r}) d{\bf r}},\\
        \end{array}
    \end{equation}
    where $\{c_i\}_{i = 1}^N$ is some set of vectors where $c_i \in {\mathbb C}^N$, $N_{e}$ is the
    total number of electrons (two in each orbital), and $s \in {\mathbb R}^{N \times N}$ is the overlap matrix.
    The optimized ground state density matrix, $\varrho_0$ from Eq.\ (\ref{Dmin}), defines the interatomic potential energy surface, $U_{\small \rm BO}({\bf R})$, within
    the Born-Oppenheimer (BO) approximation, which is given by
    \begin{equation}\label{BOP}
        {\displaystyle U_{\rm BO}({\bf R}) = F_{\rm KS}[\varrho_{0}] + 2{\rm tr}[v_{\rm ext}\varrho_0] + v_{nn}({\bf R}),}
    \end{equation}
    where $v_{nn}({\bf R})$ is the ion-ion repulsion potential. From the Born-Oppenheimer
    potential energy surface we can calculate interactomic forces that can be used in a molecular dynamics simulation.
    
    Because we are using a finite temperature ensemble with fractional occupation numbers, we are not, strictly speaking, on a Born-Opppenheimer
    potential energy surface. However, it is a straightfoward ensemble generalization of the regular Born-Oppenheimer
    potential energy surface. We will therefore still refer to the free-energy surface in Eq.\ (\ref{BOP}), which is determined by the fully relaxed (or the thermally equilibrated) electron density, as a Born-Oppenheimer potential.
    
    The constrained density-matrix minimization for $\varrho_0$ in Eqs.\ (\ref{Dmin}) and (\ref{Cnstr}) 
    is given from the solution of the nonlinear Kohn-Sham eigenvalue equation,
    \begin{equation}\label{KSEig}\begin{array}{l}
            {\displaystyle h[\varrho]c_i = \epsilon_i s c_i },
        \end{array}
    \end{equation}
    with the fractional occupation numbers given by
    \begin{equation}\label{Fermi}\begin{array}{l}
            {\displaystyle f_i  = \left(e^{\beta(\epsilon_i - \mu )}+1 \right)^{-1}. }
        \end{array}
    \end{equation}
    Where $\epsilon_i$ and $\mu$ are the molecular orbital (MO) energies and chemical potential, respectively. 
    In the Kohn-Sham Hamiltonian,
    \begin{equation}\begin{array}{l}
            {\displaystyle h[\varrho] = t_{\rm s} + v_{\rm h}[\varrho] + v_{\rm xc}[\varrho] + v_{\rm ext} },
        \end{array}
    \end{equation}
    the Hartree potential matrix, $v_{\rm h}[\varrho]$, is given by Eq.\ (\ref{VCoulomb}) and 
    and the exchange-correlation matrix, $v_{\rm xc}[\varrho]$, has matrix elements,
    \begin{equation}
        {\displaystyle \{v_{\rm xc}[\varrho]\}_{ij}
            = \frac{1}{2}\frac{\partial E_{\rm xc}[\varrho]}{\partial \varrho_{ij}}. }
    \end{equation}
    Because of the nonlinearity of the Kohn-Sham eigenvalue equation, where $\varrho$ is given by the eigenvectors in Eq.\ (\ref{Cnstr}),
    the optimized ground state solution, $\varrho_0$, is found through an iterative solution of the Kohn-Sham eigenvalue equation.
    In this optimziation procedure the Kohn-Sham Hamiltonian, $h[\varrho]$, is constructed from a mixture of previous density
    matrices that are given from the eigenvectors of previous Kohn-Sham Hamiltonians, until a stationary, self-consistent field (SCF)
    solution, is reached. This is an expensive procedure that in practice never is complete. The solution, $\varrho_0$, is therefore 
    always approximate.
    
    \section{KS-DFT+U}
    
    KS-DFT in combination with the local density or generalized gradient approximations for the exchange-correlation energy
    is an effective single-particle theory. The theory provides a computationally efficient method to calculate the physical properties
    of a broad range of materials with predictive accuracy. Nevertheless, it has some shortcomings. The main source of
    errors are the self-interaction errors and electron correlation effects for localized states \cite{JPerdew81,MPederson14,HKulik15}.
    These errors can be reduced by including orbital-dependent corrections to the Kohn-Sham matrix function where individual Kohn-Sham states are shifted in their energy levels. The orbital-dependent corrections can be derived either from Kohn-Sham DFT with self-interaction corrections or from many-particle model Hamiltonians. Here we chose to include the orbital-dependent corrections through the second approach with a KS-DFT+U correction term based on the Hubbard model \cite{VIAnisimov91,VIAnisimov97,AILichtenstein95,SLDudarev98,HKulik15}. Our orbital-corrected Kohn-Sham+U matrix function is defined by
    \begin{equation}\label{KS_U}\begin{array}{l}
            F_{\rm KS+U}[\varrho] \equiv F_{\rm KS}[\varrho] + 2{\rm tr}[u(\varrho s - \varrho s \varrho s)],\\
        \end{array}
    \end{equation}
    where $u$ is a diagonal matrix with matrix elements that can be
    tuned with respect to the different atomic orbital projections of the molecular-orbital eigenstates.
    Our $u$-dependent term is directly based one of the most commonly used DFT+U correction terms \cite{SLDudarev98}, which is translational and rotational invariant and 
    well suited for molecular dynamics simulations. 
    The $u$-dependent correction term typically also includes a spin-dependent term and has a factor $1/2$ in front, 
    which here has been replaced by a factor of $2$ for consistency with the other energy terms.
    We will only use the $u$-dependent term in Eq.\ (\ref{KS_U}) as a semi-empirical adjustment for
    materials with strong electron correlation without any particular physical interpretation of the values of $u$.
    By tuning the parameters in $u$ we simply introduce orbital-dependent
    corrections that capture some of the effects of strong electron correlation that are beyond the reach of 
    the local density or generalized gradient approximations in KS-DFT. As we will demonstrate in the simulation below, the main effect of the DFT+U correction is to adjust the electronic energy gap between the occupied and the unoccupied states. 
    
    The electronic ground state solution for KS-DFT+U is found in the same way as before using 
    the density-matrix minimization in Eq.\ (\ref{Dmin}) with the same density matrix
    constraints, $\varrho\in {\rm C}$ in Eq.\ (\ref{Cnstr}), as before, i.e.
    \begin{equation}\label{Dmin_U}
        {\displaystyle \varrho_{\rm 0} = \arg \min_{\varrho \in {\rm C}}\left\{ F_{\rm KS+U}[\varrho] + 2{\rm tr}[v_{\rm ext} \varrho] \right\}}.
    \end{equation}
    The solution to the constrained minimization is given through the same nonlinear Kohn-Sham eigenvalue problem as before, Eq.\ (\ref{KSEig}), but with the
    modified $u$-dependent effective single-particle KS-DFT+U Hamiltonian,
    \begin{equation}\label{KSH_U}\begin{array}{l}
            {\displaystyle  h[\varrho] = t_{\rm s} + v_{\rm h}[\varrho] + v_{\rm xc}[\varrho] + v_{\rm ext}}\\
            ~~\\
            {\displaystyle  
                + \frac{1}{2}(us - s\varrho us - us\varrho s + h.c.) 
            }.
        \end{array}
    \end{equation}
    The ground-state Born-Oppenheimer potential energy surface for the KS-DFT+U corrected Kohn-Sham matrix function is then given by
    \begin{equation}
        {\displaystyle  U_{\rm BO+U}({\bf R}) = F_{\rm KS+U}[\varrho_{\rm 0}] + 2{\rm tr}[v_{\rm ext}\varrho_{\rm 0}] + v_{nn}({\bf R})}.
    \end{equation}
    This potential can then be used to calculate the interatomic forces and integrate the equations of motion,
    \begin{equation}
        M_I {\bf \ddot R}_I = - \nabla_I U_{\rm BO+U}({\bf R})
    \end{equation}
    in a molecular dynamics simulation, where $\{M_I\}$ are the atomic masses.

    \section{XL-BOMD with DFT+U}
    
    The main cost of a QMD simulation based on KS-DFT is the cost of finding the (thermally)
    relaxed self-consistent ground state, $\varrho_{\rm 0}$, prior to the force evaluation in each time step.
    The iterative solution of the nonlinear eigenvalue problem, Eq.\ (\ref{KSEig}),
    is expensive with a prefactor that scales linearly with the number
    of iterations required to find a sufficiently converged self-consistent ground-state solution.
    By using a good initial guess to the SCF optimization, which can be generated from an extrapolation of the ground state density matrix from previous time steps, it is possible to significantly reduce the computational overhead.
    However, because the iterative ground state optimization is approximate, the calculated forces are 
    never exact and there is an inconsistency between the calculated forces and the exact Born-Oppenheimer ground state 
    potential energy surface. The extrapolation in combination with an incomplete ground-state optimization leads 
    to non-conservative forces and a systematic drift in the total energy, because of
    a broken time-reversal symmetry in the fictitious propagation of the underlying electronic degrees of freedom that is generated through
    the extrapolation \cite{DRemler90,PPulay04,ANiklasson07}.  Alternatively, we may restart the ground state optimization in each new time step
    from overlapping atomic densities, which preserves the time-reversal symmetry and avoids a systematic drift in the total energy, but the computational
    cost is significantly higher.  XL-BOMD
    \cite{ANiklasson08,PSteneteg10,GZheng11,MCawkwell12,JHutter12,LLin14,MArita14,PSouvatzis14,ANiklasson14,KNomura15,AAlbaugh15,ANiklasson17,JBjorgaard18}
    is a framework that has been developed to avoid these shortcomings.
    
    XL-BOMD is based on the concept of backward error analysis or a \textit{shadow} Hamiltonian approach \cite{SToxvaerd94,GJason00,ShadowHamiltonian,SToxvaerd12}. Instead of calculating approximate forces using
    an expensive iterative ground-state optimization procedure for an underlying ``exact'' Born-Oppenheimer potential
    energy surface, we can calculate exact forces in a fast and simple way, but for an underlying approximate \textit{shadow} potential energy surface that closely follows the ``exact'' regular Born-Oppenheimer potential. In this way we can reduce the computational cost and at the same time restore a consistency between the
    calculated forces and the underlying shadow potential. With the consistent, conservative forces we can then generate stable molecular 
    trajectories at only a fraction of the cost of regular direct Born-Oppenheimer molecular dynamics simulation.

    \subsection{The Shadow Potential}
    
    In XL-BOMD a shadow free energy matrix function is constructed from a linearization of the KS-DFT+U energy function, Eq.\ (\ref{KS_U}), around some approximate solution, $\nu \in {\mathbb R}^{N\times N}$, to the exact ground state density matrix, $\varrho_{\rm 0}$ \cite{ANiklasson14,ANiklasson17,ANiklasson20b}.
    The constrained stationary minima of this shadow energy matrix functional then generates the shadow Born-Oppenheimer potential.
    The shadow matrix functional for the orbital-corrected KS-DFT+U matrix function is given by
    \begin{align}\label{KSF1_U}
        {\cal F}_{\rm KS+U}[\varrho, \nu ] &= F_{\rm KS+U}[ \nu ] 
        \nonumber\\ &
        ~~+ {\rm tr}\left[ (\varrho- \nu ) \left( \partial F_{\rm KS+U}[\varrho]/\partial \varrho \right) \vert_{\varrho={\nu }} \right] \nonumber\\
        &= 2{\rm tr}[t_{\rm s}\varrho ] + {\rm tr}[(2\varrho- \nu ) v_{\rm h}[ \nu ]] + E_{\rm xc}[ \nu ] \nonumber\\
        &~~+ 2{\rm tr}\left[(\varrho- \nu )v_{\rm xc}\left[\nu  \right]\right] + E_{\rm ent}(f)\nonumber\\
        &~~+2{\rm tr}\left[u(\varrho s- \nu s\varrho s-\varrho s\nu s+\nu s\nu s)\right].
    \end{align}
    The stationary ground-state solution, $\varrho_{\rm 0}[\nu ]$, of the linearized matrix function 
    is $\nu $-{\em dependent} and is found by a constrained density-matrix minimization with the same density matrix
    constraints, $\varrho \in {\rm C}$ as before in Eq.\ (\ref{Cnstr}), and
    \begin{equation}\label{Dmin_U}
        {\displaystyle \varrho_{\rm 0}[\nu ] = \arg \min_{\varrho \in {\rm C}}\left\{ {\cal F}_{\rm KS+U}[\varrho,{\nu }] + 2{\rm tr}[v_{\rm ext} \varrho] \right\}}.
    \end{equation}
    Because of the linearization, the minimization can be solved in a single step as a solution to
    a {\em linear} Kohn-Sham eigenvalue problem
    \begin{equation}\label{LKSEig}\begin{array}{l}
            {\displaystyle h[\nu ]c_i = \epsilon_i s c_i },
        \end{array}
    \end{equation}
    where
    \begin{equation}\label{D_P}\begin{array}{l}
            {\displaystyle \varrho_{\rm 0}[{\nu }] = \sum_i f_i c_i c_i^\dagger}
        \end{array}
    \end{equation}
    and the fractional occupation numbers given from the Fermi function in Eq.\ (\ref{Fermi}).
    The Kohn-Sham Hamiltonian of the linearized orbital-corrected KS-DFT+U matrix function is given by
    \begin{equation}\label{H_P}\begin{array}{l}
            {\displaystyle h[\nu ] = t_{\rm s} +v_{\rm h}[\nu ] + v_{\rm xc}[\nu ] + v_{\rm ext} }\\
            ~~\\
            + \frac{1}{2}(su - su\nu s - s\nu su  + h.c.).
        \end{array}
    \end{equation}
    The $\nu $-dependent shadow Born-Oppenheimer potential energy surface, ${\cal U}_{\rm BO+U}({\bf R}, \nu )$, is then given in the same way as before, but using the linearized
    KS-DFT+U matrix function,
    \begin{equation}\label{ShadowPot}\begin{array}{l}
            {\displaystyle {\cal U}_{\rm BO+U}({\bf R}, \nu ) = {\cal F}_{\rm KS+U}[\varrho_{\rm 0}[\nu ],\nu ]}\\
            ~~\\
            {\displaystyle ~ + 2{\rm tr}\left[\varrho_{\rm 0}[\nu ]\times v_{\rm ext}\right] + v_{nn}({\bf R}).}
        \end{array}
    \end{equation}
    
    The difference between the shadow potential energy surface, ${\cal U}_{\rm BO+U}({\bf R},\nu )$, 
    and the ``exact'' fully converged Born-Oppenheimer potential energy surface, $U_{\rm BO+U}({\bf R})$, is small
    if the residual matrix function, $\varrho_{\rm 0}[\nu ]  - \nu $,
    is small. The difference scales as ${\cal O}(\| \varrho_{\rm 0}[\nu ]  - \nu  \|^2)$.
    
    The $\nu $-dependent approximate ground state, $\rho_0[\nu ]$, is different from the exact ground state density matrix, $\rho_0$,
    of the exact fully converged Born-Oppenheimer potential, but $\rho_0[\nu ]$ is still the exact fully converegd ground state
    solution of the shadow potential. The first-order variation of the shadow potential with respect to the density matrix
    around $\rho_0[\nu ]$ therefore vanish, i.e.\ $\partial {\cal U}_{\rm BO+U}/\partial \rho \vert_{\rho = \rho_0[\nu ]} = 0$.
    This is important in the calculation of the interatomic forces, because it means that the partial force term including 
    $\left(\partial {\cal U}_{\rm BO+U}/\partial \rho \vert_{\rho = \rho_0[\nu ]}\right)\left(\partial \rho/\partial {\bf R}_I\right) = 0$ will vanish,
    which simplifies the calculation of the forces, without relying on the Hellmann-Feynman theorem or additional adjustment terms.

    \subsection{Extended Lagrangian}

    In a molecular dynamics simulation the atoms are moving and at some point the approximate ground state density matrix, $\nu $,
    around which we performed the linearization of the KS-DFT+U matrix energyfunction in Eq.\ (\ref{KSF1_U}) will no longer be close to the exact 
    ground state $\varrho_0$.
    We therefore need to update $\nu $ along the molecular trajectory to keep it close to the unknown ground state $\varrho_0$.
    Without an update the linearization of the KS-DFT+U matrix function, ${\cal F}_{\rm KS+U}[\varrho,\nu ]$, will eventually deteriorate and 
    the difference between the shadow potential and the fully converged ``exact'' Born-Oppenheimer potential energy surfaces may diverge. To simply update
    $\nu $ with the atomic positions, ${\bf R}$, would require the calculation of $\partial \nu /\partial {\bf R}_I$ terms, and their effect
    on the $\nu $-dependent potential energy surface. In general, this would be quite expensive. Instead, in XL-BOMD the approximate ground state density matrix, $\nu $, is included
    as a dynamical tensor variable that evolves through a harmonic oscillator that is centered around the ground state, $\varrho_{\rm 0}$,
    or the best available approximation, which in our case is $\varrho_{\rm 0}[\nu ]$. The dynamics is defined through the extended Lagrangian,
    \begin{equation}\label{XL}\begin{array}{l}
            {\displaystyle {\cal L}({\bf R}, {\bf \dot R}, \nu , {\dot \nu }) =
                \frac{1}{2} \sum_I M_l{\dot R}_I^2 - {\cal U}_{\rm BO+U}({\bf R},\nu )} \\
            ~~\\
            {\displaystyle ~ + \frac{\mu}{2} {\rm tr}[{\dot \nu }^2]
                - \frac{\mu \omega^2}{2} {\rm tr}[(\varrho_{\rm 0}[\nu ] - \nu )^T{\cal T}(\varrho_0[\nu ]-  \nu )]}.
        \end{array}
    \end{equation}
    Here ${\bf R}$ and ${\bf {\dot R}}$ are the atomic positions and their velocities; $\nu $
    and ${\dot \nu }$ are the dynamical matrix variables that represent the extended electronic degrees of freedom; 
    ${\cal U}_{\rm BO+U}({\bf R}, \nu )$ is the shadow potential for the electronic free energy based on the linearized KS-DFT+U matrix function
    at some electronic temperature, $T_e \ge 0$, that approximates the corresponding exact Born-Oppenheimer potential energy surface; 
    ${\cal T} \equiv {\cal K}^T{\cal K}$ is a symmetric positive definite metric tensor of the harmonic well that makes $n$ oscillate around an even closer
    approximation to the exact ground state than $\varrho_0[\nu ]$ and will be defined below;
    $\mu$ is a fictitious electronic mass parameter; and $\omega$ is the frequency of the harmonic oscillator extension that
    defines the time scale for the dynamics of the extended electronic degrees of freedom.
    We may use different representations of the extended electronic degrees of freedom $\nu $.  Instead of the atomic-orbital matrix
    representation, $\nu $, we can use an orthogonal representation, $\nu ^\perp = z^{-1} \nu  z^{-T}$ where $z$ is chosen such that 
    $\sum_{kl} z_{ki}s_{kl}z_{lj} = \delta_{ij}$, or we can chose a modified dynamical variable $x = \nu s$.
    For simplicity, we will here express the dynamics in terms of the atomic-orbital representation, $\nu $, but it is straightforward to use also the other
    representations. The choice of dynamical variables, $x \equiv \nu s$ and ${\dot x},$ as in Ref.\ \cite{MArita14} seems to be slightly more efficient and is a more natural choice because of its consistent tensorial behavior under integration. This is also the version that we will use in the examples demonstrating XL-BOMD using a DFT+U functional in section \ref{Examples}.
    
    The expression for the harmonic oscillator of the extended Lagrangian in Eq.\ (\ref{XL}) includes a metric tensor,
    \begin{equation}\label{Tensor}
        {\displaystyle {\cal T} \equiv {\cal K}^T{\cal K} },
    \end{equation}
    where ${\cal K}$ is a kernel that acts as a fourth-order tensor, which performs mappings between matrices.
    This kernel, ${\cal K}$, is defined from the inverse of the Jacobian, ${\cal J}$, of the residual matrix function, where
    \begin{equation}\label{Jacobian}
        {\displaystyle  {\cal J}_{ij,kl}  = \frac{\partial (\{{\varrho_{\rm 0}[\nu ]}\}_{ij}-\nu _{ij})}{\partial \nu _{kl}}, }
    \end{equation}
    and
    \begin{equation}\label{InverseJacobian}
        {\displaystyle  {\cal K} = {\cal J}^{-1}. }
    \end{equation}

    \subsection{Equations of motion}
    
    The atomic coordinates typically evolve on a slow time scale compared to the electronic motion.
    If initially the electrons are in the ground state we may therefore assume they will evolve 
    close to the electronic ground state as the atoms are moving. This adiabatic assumption 
    is the reasoning behind the Born-Oppenheimer approximation in
    quantum-based molecular dynamics simulations \cite{WHeitler27,MBorn27,DMarx00}.
    In the derivation of the equations of motion of XL-BOMD from Euler-Lagrange's equations 
    we can also apply an adiabatic approximation that separates the motion between the nuclear and the extended electronic degrees of freedom.
    Our derivation of the equations of motion of XL-BOMD are therefore performed in 
    in an adiabatic limit where $\omega \rightarrow \infty$ and $\mu \rightarrow 0$ 
    such that $\mu \omega = {\rm constant}$. This is a classical analogue to the Born-Oppenheimer approximation, where
    the extended electronic degrees of freedom is assumed to evovle on a fast time scale compared to the motion of the
    atomic positions \cite{ANiklasson14}. In this adiabatic limit we get the equations of motion
    \begin{equation}\label{EqR}
        {\displaystyle  M_I {\ddot R}_I = - \left. \nabla_I {\cal U}_{\rm BO+U}({\bf R},\nu ) \right \vert_\nu ,   }
    \end{equation}
    for the nuclear degrees of freedom and
    \begin{equation}\label{EqX}
        {\displaystyle  {\ddot \nu } = - \omega^2 {\cal K} \left(\varrho_{\rm 0}[\nu ] - \nu \right), }
    \end{equation}
    for the electronic degrees of freedom.
    The corresponding constant of motion is given by the total energy,
    \begin{equation}\label{Constant}
        {\displaystyle  { E}^{\rm tot}_{\rm BO+U} = \frac{1}{2}\sum_I M_I \vert {\bf \dot R}_I\vert^2 +  {\cal U}_{\rm BO+U}({\bf R},\nu )}.
    \end{equation} 
    These are the central equations of XL-BOMD, which are exact in continuous time, and can be used to generate the molecular trajectories in QMD simulations.
    
    In the adiabatic limit the residual function $\|\left(\varrho_{\rm 0}[\nu ] - \nu \right)\| \propto \omega^{-2}$, which simplifies
    the evaluation of the iteratomic forces in the first equation, Eq.\ (\ref{EqR}).
    We can express the equations of motion in Eq.\ (\ref{EqR}) as
    \begin{equation}\label{Eq_R}\begin{array}{l}
            {\displaystyle  M_I {\bf \ddot R}_I = - \left. \frac{\partial {\cal U}_{\rm BO+U}({\bf R},\nu )}{\partial {\bf R}_I}\right \vert_\nu  \equiv - {\cal U}'_{\rm BO+U}({\bf R},\nu ),   }\\
            ~~\\
            {\displaystyle  M_I {\bf \ddot R}_I = -2{\rm tr}[t_s'\varrho_0[\nu ] + {\rm tr}[(2\varrho_0[\nu ]-\nu )v'_{\rm h}[\nu ] }\\
            ~~\\
            {\displaystyle  ~~~~~ + E_{\rm xc}'[\nu ] + 2{\rm tr}[(\varrho_0[\nu ] -\nu )v'_{\rm xc}[\nu ]]  }\\
            ~~\\
            {\displaystyle  ~~~~~ + 2 {\rm tr}[u(\varrho_0[\nu ]s' - \nu s'\varrho_0[\nu ]s - \nu s\varrho_0[\nu ]s')]}\\
            ~~\\
            {\displaystyle  ~~~~~ + 2 {\rm tr} [v'_{\rm ext}\varrho_0[\nu ]] + v'_{\rm nn}({\bf R})},
        \end{array}
    \end{equation}
    where we use the prime notation, $'$, for the partial derivative with respect 
    to the nuclear coordinates under constant $\nu$, e.g.\ ${\cal U}' \equiv \partial {\cal U}/\partial {\bf R}_I\vert_\nu$.
    
    The forces above are the exact conservative forces for the shadow Born-Oppenheimer potential. Because $\delta {\cal U}_{\rm BO+U}({\bf R},\nu ) \big / \delta \varrho = 0$ at $\varrho = \rho_0[\nu]$, any force terms with $\partial \varrho/\partial {\bf R}_I$ can be ignored. The force expression we use above therefore has the same simplicity as a Hellman-Feynman force expression. Here this is possible even if $\rho_0[\nu]$ is not the exact regular ground state. 
    
    The shadow Born-Oppenheimer potential, ${\cal U}_{\rm BO+U}({\bf R},\nu )$ in Eq.\ (\ref{ShadowPot}), can be seen as a generalized Harris-Foulkes functional \cite{JHarris85,MFoulkes89} for orbital-dependent Kohn-Sham corrections. However, because ${\cal U}_{\rm BO+U}({\bf R},\nu )$ is given as a variationally optimized ground state of a shadow matrix energy function, and as $\nu$ appears as a dynamical variable within the extended Lagrangian formulations, no partial derivatives, $\partial \nu \big / \partial {\bf R}_I$, appear in the force expression.  In contrast to a Harris-Foulkes expression we can therefore calculate forces and these forces are exact for ${\cal U}_{\rm BO+U}({\bf R},\nu )$.
    
    The kernel ${\cal K}$ in Eq.\ (\ref{EqX}) is defined as the inverse Jacobian of the residual in Eqs.\ (\ref{InverseJacobian}) and (\ref{Jacobian})
    and therefore acts as a Newton step in an iterative solution of a system of nonlinear equations, 
    i.e.\ the residual matrix function equation $\varrho_{\rm 0}[\nu ] - \nu  = 0$. 
    The dynamical matrix $\nu$ therefore behaves as if it would oscillate around a much closer approximation to the
    exact ground state, $\varrho_{\rm 0}$, than $\varrho_{\rm 0}[\nu ]$, because
    \begin{equation}\label{EqK} 
        {\displaystyle  {\cal K} \left(\varrho_{\rm 0}[\nu ] - \nu \right)  =  \left(\varrho_{\rm 0} - \nu \right)  + {\cal O}\left(\|\varrho_{\rm 0}[\nu ] - \nu \|^2\right).}
    \end{equation}
    Unfortunately, it is expensive to calculate the exact kernel and instead we need to use some approximation
    in the integration of the electronic degrees of freedom. Either a scaled delta function, ${\cal K} \approx - c {\cal I}$,
    with $c \in [0,1]$ can be used or a more accurate low-rank Krylov subspace approximation, which we will present below.

    \subsection{Integrating the equations of motion}
    
    To integrate the equations of motion, Eqs.\ (\ref{EqR}) and (\ref{EqX}), a modified leapfrog velocity Verlet scheme can be used \cite{ANiklasson09,PSteneteg10,GZheng11}, which includes an additional dissipative term in the integration of the extended electronic degrees of freedom. This additional term breaks the time-reversal symmetry to
    some chosen higher odd-order in the integration time step, $\delta t$, which dampens the accumulation of numerical noise
    that otherwise could cause instabilities in a perfectly reversible integration. In this
    way the evolution of the electronic degrees of freedom stays synchronized to the dynamics of the nuclear motion.
    The modified leapfrog velocity Verlet integration scheme for the integration
    of the nuclear and electronic degrees of freedom is given by
    \begin{equation}\label{Integration}\begin{array}{l}
            {\displaystyle {{\dot {\bf R}}}(t + \frac{\delta t}{2})  = {{\dot {\bf R}}}(t) + \frac{\delta t}{2} {{\ddot {\bf R}}}(t)},\\
            ~~\\
            {\displaystyle {{\bf R}}(t + \delta t) = {{\bf R}}(t) + \delta t {{\dot {\bf R}}}(t + \frac{\delta t}{2})},\\
            ~~\\
            {\displaystyle \nu (t + \delta t) = 2\nu (t) - \nu (t - \delta t) + \delta t^2 {\ddot \nu }(t)}\\
            
            {\displaystyle ~~~~~~~~~~~~~~~~~ + \alpha \sum_{k = 0}^{k_{\rm max}} c_k \nu (t-k \delta t)},\\
            ~~\\
            {\displaystyle {{\dot {\bf R}}}(t + \delta t) = {{\dot {\bf R}}}(t + \frac{\delta t}{2})
                + \frac{\delta t}{2} {{\ddot {\bf R}}}(t+\delta t)}.\\
        \end{array}
    \end{equation}
    The last term in the integration of $\nu (t)$ is the additional damping term,
    where the coefficients, $\alpha$ and $\{ c_k \}_{k = 0}^{k_{\rm max}}$,
    as well as a dimensionless constant, $\kappa = \delta t^2 \omega^2$,
    have been optimized for various values of $k_{\rm max}$ and are given in Ref.\ \cite{ANiklasson09}.
    
    In the initial time step $\nu (t_0-k\delta t)$ for $k = 0,1,\ldots,k_{\rm max}$ are all set to the
    fully converged regular Born-Oppenheimer ground state density, $\varrho_{\rm 0}$, i.e.\
    at $t_0$ we set $\nu(t_0-k\delta t) = \varrho_{\rm 0}$ for $k = 0,1,\ldots,k_{\rm max}$.
    A reasonably well-converged iterative self-consistent field optimization is thus required,
    but only in the first initial time step.  
    The modified Verlet integration scheme above works well without any significant drift in the constant of motion on time scales
    relevant for quantum-based Born-Oppenheimer molecular dynamics.
    Several alternative integration schemes for XL-BOMD have also
    been proposed and analyzed \cite{AOdell09,AOdell11,AAlbaugh15,VVitale17,AAlbaugh17,AAlbaugh18}, but will not be used in this article.
    
    \subsection{Krylov subspace approximation of the kernel} 
    
    A key challenge in the integration of the electronic degrees of freedom, Eq.\ (\ref{Integration}),
    is the calculation of ${\ddot \nu}(t)$, which is given by Eq.\ (\ref{EqX}). By using a low-rank
    Krylov-subspace approximation \cite{ANiklasson20}
    of the kernel, ${\cal K}$, adapted to the density matrix formalism \cite{ANiklasson20b,Maksim23}, we can approximate ${\ddot \nu}(t)$ as
    \begin{equation}\label{LowRank}\begin{array}{l}
            {\displaystyle  {\ddot \nu} = - \omega^2 {\cal K} \left(\varrho_{\rm 0}[\nu] - \nu\right) }\\
            ~~\\
            {\displaystyle \approx - \omega^2  \sum_{i,j = 1}^m v_i g_{ij} \langle w_j, (\varrho_{\rm 0}[\nu] - \nu)\rangle  }.
        \end{array}
    \end{equation}
    The matrices $v_i$, $w_i \in {\mathbb R}^{N \times N}$ and $g \in {\mathbb R}^{m\times m}$,
    are based on a rank-$m$ Krylov subspace approximation and are generated through Algorithm 1.
    We use the matrix inner product notation, $\langle v_i, v_j \rangle = {\rm tr}[v_i^T v_j]$.
    The algorithm requires the calculation of the perturbation in the density matrix, $\partial \varrho_{\rm 0}[\nu+\lambda v_m]/\partial \lambda$ at $(\lambda = 0)$. 
    This density matrix response can be calculated through the intermediate perturbation to first order in $\lambda$ in the Kohn-Sham Hamiltonian, i.e.\
    $h_0 + \lambda h_1 \approx h[\nu + \lambda v_m]$, which can be performed with regular 
    Rayleigh-Schr\"{o}dinger perturbation theory when $T_e = 0$, or
    for fractional occupation numbers when $T_e > 0$ \cite{SLAdler62,NWiser63,ANiklasson15,YNishimoto17,ANiklasson20,ANiklasson20b}. The pseudocode of computing Krylov subspace approximation of the kernel is shown in Algorithm~\ref{KernelApproximation}.

   \begin{algorithm}[H]
         \caption{
            \label{KernelApproximation}
            This algorithm generates the matrices $\{v_i\}$, $g$, and $\{ w_i\}$ for the kernel approximation, Eq.\ \ref{LowRank},
            in the integration of the electronic degrees of freedom, Eqs.\ \ref{EqX} and \ref{Integration}, using
            a rank-$m$ Krylov subspace approximation of the inverse Jacobian kernel \cite{ANiklasson20,ANiklasson20b}, where
            ${\cal K}(\varrho_{\rm 0}[\nu] - \nu) \approx \sum_{i,j}^m v_i g_{ij} \langle w_j, (\varrho_{\rm 0}[\nu] - \nu)\rangle$.
            The inner product is given by $\langle v_i, v_j \rangle = {\rm Tr}[v_i^T v_j]$, 
            with the Frobenius matrix norm $\| v_i\| = \sqrt{\langle v_i, v_i \rangle}$.
            The trace conserving canonical density matrix perturbation, 
            $\left. \partial_\lambda  \varrho[\nu+\lambda v_m] \right \vert _{\lambda = 0}$, can be
            performed as in Ref.\ \cite{ANiklasson15,ANiklasson20,ANiklasson20b}.
         }
         \algsetup{indent=1em}
       \begin{algorithmic}
              \STATE $ w_0 = (\varrho_{\rm 0}[\nu] - \nu), ~~ m = 0$
            \WHILE{Error  $ >$ Chosen Tolerance}
            \STATE $ m = m + 1$
            \STATE $  v_m =  w_{m-1}$
            \STATE $  v_m =  v_m - \sum_{j = 1}^{m-1}\langle v_i, v_j \rangle v_j$
            \STATE $  v_m =  v_m \|v_m\|^{-1}$
            \STATE $  w_{m} =  \left( \left. \partial_\lambda \varrho[\nu+\lambda v_m] \right \vert _{\lambda = 0} - v_m\right)$
            \STATE $ o_{ij} = \langle w_i, w_j \rangle, ~i,j = 1,2,\ldots, m$
            \STATE $ g = o^{-1} $
            \STATE Error = $\left \|\left(\sum_{i,j = 1}^m w_i g_{ij} \langle w_j, w_0\rangle - w_0\right)\right \|/\|w_0\| $
            \ENDWHILE
           \STATE $\Rightarrow {\cal K}\left(\varrho_{\rm 0}[\nu] - \nu\right) 
            \approx \sum_{i,j}^m v_i g_{ij} \langle w_j,(\varrho_{\rm 0}[\nu] - \nu)\rangle$
    \end{algorithmic}
  \end{algorithm}
    
    \section{Examples}\label{Examples}
    
    To demonstrate the QMD simulation framework combining KS-DFT+U and XL-BOMD, we will first look at a simple molecular system of liquid nitromethane, CH$_3$NO$_2$, to demonstrate the ability to tune the electronic HOMO-LUMO energy gap, while the rest of the dynamics behave in the same way. We then apply the combined scheme using XL-BOMD with DFT+U to simulations of nuclear fuel, UO$_2$. This is a well-known example where regular KS-DFT fails to capture some of the most important electronic structure properties. In particular, the electronic gap is missing and KS-DFT calculations show a metallic behavior of UO$_2$, whereas the physically correct picture has an electronic gap of about 2 eV \cite{HIdriss10,HHeming13}.
    
    In our electronic structure theory we will use an approximate KS-DFT+U scheme based on self-consistent charge density-functional based tight-binding (SCC-DFTB) theory 
    \cite{JHarris85,MFoulkes89,DPorezag95,MElstner98,MFinnis98,TFrauenheim00,MGaus11,BAradi15,BHourahine20} as implemented in the LATTE electronic structure package \cite{LATTE, MCawkwell12,AKrishnapriyan17}. 
    SCC-DFTB has previously been developed to include KS-DFT+U corrections \cite{BHourahine20,BHourahine07,SSanna08} 
    as well as XL-BOMD \cite{GZheng11,MCawkwell12,BAradi15}, but only separately. In the combined KS-DFT+U and XL-BOMD framework applied here, we will use the density matrix times the overlap matrix, $x = \nu s$, and its time derivative, ${\dot x}$, as our dynamical field variables \cite{MArita14,ANiklasson20b,Maksim23} instead of the density matrix itself. This density matrix formalism for XL-BOMD is presented in Ref.\ \cite{ANiklasson20b} including the low-rank Krylov subspace approximation, Eq.\ (\ref{LowRank}), for the integration of equation of motion in Eq.\ (\ref{EqX}) with $x = \nu s$ and ${\dot x}$ as the dynamical matrix variables. What are different in our simulation examples is the shadow energy functional, ${\cal F}_{\rm KS+U}[\varrho, \nu ]$ in Eq.\ (\ref{KSF1_U}), the corresponding shadow potential, ${\cal U}_{\rm BO+U}({\bf R},\nu)$ in Eq.\ (\ref{ShadowPot}), and the force term in Eq.\ (\ref{Eq_R}).
    
    \subsection{Nitromethane}
    
\begin{figure}[t]\centering 
        \resizebox*{3.9in}{!}{\includegraphics[angle=00]{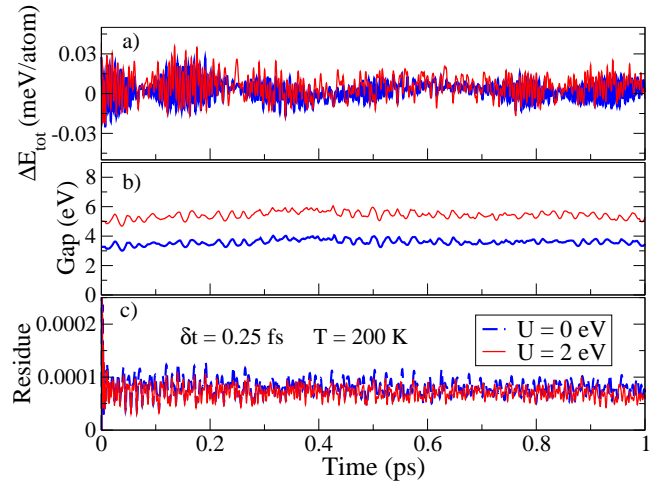}}
        \caption{\label{NM_DFTB_U}\small Combined SCC-DFTB+U XL-BOMD NVE simulation (with periodic boundary conditions) of liquid nitromethane (CH$_3$NO$_2$)$_7$ with a Hubbard U set to 0 eV or 2 eV. The upper panel a) shows the fluctuations in the total energy (potential + kinetic) around the average. The statistical temperature was around 200 K with an integration time step $\delta t = 0.25$ fs. The residue in the lower panel c) was given by the Frobenius norm of the density matrix residual function, $\|\varrho[\nu]s - \nu s \|_{\rm F}$. The mid panel b) shows the fluctuation of the electronic HOMO-LUMO energy gap (Gap).}
    \end{figure}

\begin{figure}[t]\centering 
        \resizebox*{3.7in}{!}{\includegraphics[angle=00]{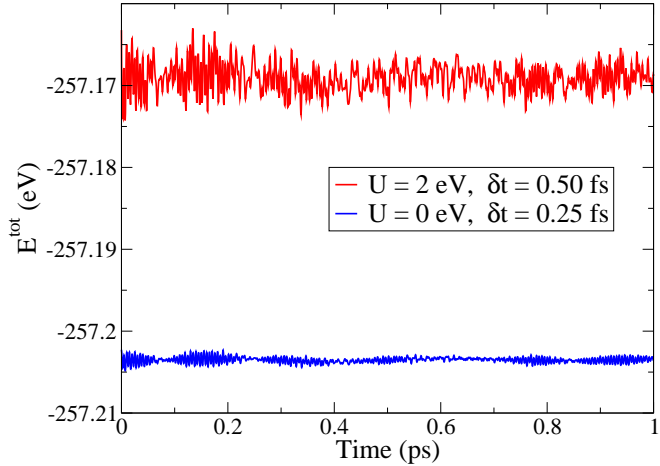}}
        \caption{\label{NM_DFTB_U_2}\small 
            Combined SCC-DFTB+U XL-BOMD NVE simulations of liquid nitromethane (CH$_3$NO$_2$)$_7$ with a Hubbard U set to 0 eV or 2 eV for a time step of $\delta t = 0.25$ fs and $\delta t = 0.50$ fs. The total energy is shifted upwards for U = 2 eV and the amplitude of the fluctuations increases approximately by a factor of 4 as the size of the integration time step is increased by a factor of 2. The statistical temperature fluctuated around 200 K.}
    \end{figure}

    Figure \ref{NM_DFTB_U} shows a combined SCC-DFTB+U and XL-BOMD microcanonical (NVE) simulation of liquid nitromethane, (CH$_3$NO$_2$)$_7$, where the Hubbard U parameter is set 0 eV or 2 eV. The fluctuations in the total energy around their average value and the residue given by the Frobenius norm of the density matrix residual function, $\|\varrho[\nu]s - \nu s\|_{\rm F} = \|\varrho[\nu]s - x\|_{\rm F}$, are following each other closely for the two cases, as is shown in the upper panel a) and lower panel c). The main difference is the size of the electronic HOMO-LUMO energy gap shown in the mid panel b). The only difference is a shift of about 2 eV. The total energy remains stable with no visible drift in the total energy. While the fluctuations in the total energy behave in the same way around their average values, the total energy is shifted. This is seen in Fig.\ \ref{NM_DFTB_U_2} where an increased Hubbard U leads to a shift in the total energy. In this figure we also see how the amplitude of the total energy fluctuations for the Verlet integration scheme scales approximately as $\delta t^2$, i.e.\ the amplitude is increased by a factor of 4 as we double the size of the integration time step from $\delta t = 0.25$ fs to  $\delta t = 0.50$ fs. Also the size of the residual error, $\|\varrho[\nu]s - \nu s\|_{\rm F}$, which provides a measure of the difference to the exact regular ground state solution, scales quadratically with the integration time step (not shown). The error in the potential energy surface scales with the square of the residual, i.e., $\| {\cal U}_{\rm BO+U}({\bf R},\nu) - U_{\rm BO+U}({\bf R})\| \propto \|\varrho[\nu]s  - \nu s \|^2$, and the error in the sampling of the potential energy surface therefore scales as $\delta t ^4$ \cite{ANiklasson20b,ANiklasson21b}.
    
    This example demonstrates the ability of the combined KS-DFT+U and XL-BOMD simulation scheme to alter the size of the HOMO-LUMO gap, while providing stable molecular trajectories. The ability to tune the gap can be of significant importance if we need to modify the response properties of a material. For approximate DFT methods like SCC-DFTB or semi-empirical quantum-chemistry methods \cite{MDewar77,MDewar85,JStewart13,PDral15,Anatole15,DYaron18,JKranz18,CBannwarth18,NGoldman18,PDral19,WMalone20,CBannwarth20,ZGuoqing20,PZheng21,ZGuoqing22,DYaron18}, the molecular polarizability, which may affect the long-range Coulomb interactions between polarized molecules, could be tuned by modifying the Hubbard-U parameter using the DFT+U correction.

    \subsection{UO$_2$}
    
    \begin{figure}[t]\centering
        \resizebox*{3.5in}{!}{\includegraphics[angle=00]{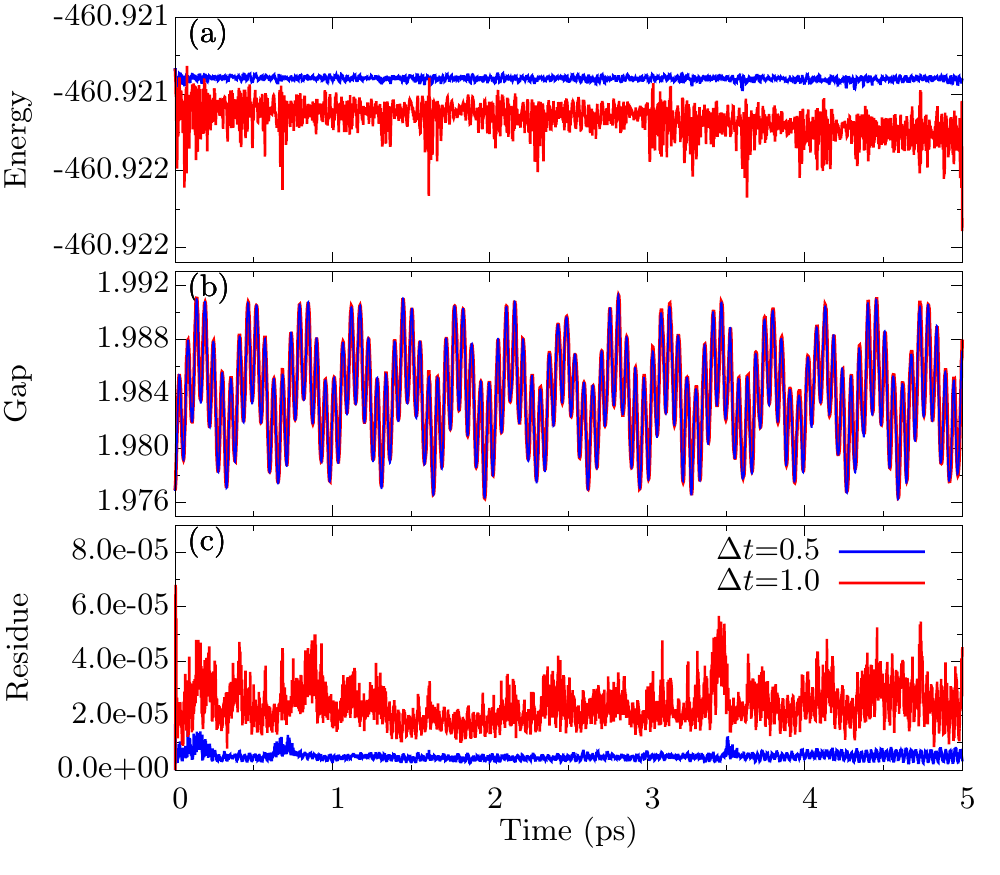}}
        \caption{\label{UO2_DFTB_U}\small DFTB+U XL-BOMD based NVE simulation of a 96 atom UO$_2$ supercell using a Hubbard U = 2 eV. The upper panel a) shows the shifted fluctuations in the total energy and the mid panel the size of the electronic HOMO-LUMO gap. The residue in the lower panel c) was given by the Frobenius norm of the density matrix residual function, $\|\varrho[\nu]s - \nu s\|_{\rm F}$. Two different integration time steps were used}
    \end{figure}

    Figure \ref{UO2_DFTB_U} illustrates the outcomes of microcanonical (NVE) simulations of a 96-atom supercell (periodic boundary conditions) of nuclear fuel, UO$_2$, employing the combined SCC-DFTB+U and XL-BOMD simulation approach with a Hubbard U = 2 eV. The simulations are performed for two different time steps, $\delta t = 0.25$ fs and $\delta t = 0.5$ fs. Without a Hubbard-U parameter, UO$_2$ is metallic, lacking an electronic energy gap, and does not match experimental observations, where a gap of about 2 eV is seen.
    The SCC-DFTB+U parameterization relies on first principled KS-DFT calculations fitted to first principles calculations \cite{Cawkwell_UO2}.
    The top panel presents fluctuations in the total energy that are shifted such that the initial total energy is set to 0. There is no visible systematic drift, and we observe an approximate $\delta t^2$ scaling, i.e.\ the amplitude increases by a factor of 4 as the time step is doubled in size. The middle panel displays the size of the electronic energy gap, which oscillates near 2 eV, close to the chosen Hubbard-U value.
    The bottom panel depicts the Frobenious norm of the matrix residual function is on the order of $10^{-5}$. This residue represents the difference to the exact ground state solution equivalent to a self-consistency error in a regular Born-Oppenheimer simulation. As the integration time step is halved, the size of the residual is reduced by a factor of 4, demonstrating the approximate $\delta t^2$ scaling of the residual error. As discussed above, this gives an error in the potential energy surface that scales as $\delta t^4$ \cite{ANiklasson17,ANiklasson21b}.

    \section{Summary and discussion}
    
    We have presented a framework for QMD simulations that combines DFT+U and XL-BOMD. In this way we have been able to reduce the computational cost of QMD simulations also for systems with electron correlation effects beyond the reach of regular KS-DFT based on the local density or generalized gradient approximations. With the extended Lagrangian formulation this is achieved without requiring an iterative self-consistent-field optimization of the electronic ground state prior to the force evaluations, which is necessary in regular direct Born-Oppenheimer molecular dynamics simulations. The method provides accurate and stable molecular trajectories at the same time as the computational cost per time step is drastically reduced by avoiding the iterative SCF optimization that normally is required prior to each force evaluation in a regular Born-Oppenheimer simulation.
    
    The basic idea behind our approach can be traced back to a backward error analysis or a shadow Hamiltonian approach \cite{SToxvaerd94,GJason00,SToxvaerd12,KDHammonds20,ShadowHamiltonian}. This is a conceptually simple but highly powerful idea.
    Instead of calculating {\em approximate} solutions for an underlying {\em exact regular} Born-Oppenheimer potential, we do the opposite. Instead, we calculate the {\em exact} electron density, energies, and forces, but for an underlying {\em approximate shadow} Born-Oppenheimer potential. In this way the calculated forces are conservative with respect to the approximate shadow potential and generate accurate molecular trajectories with long-term energy stability. Here we have shown how this concept can be extended beyond regular KS-DFT to include also orbital-dependent DFT+U corrections.
    
    Our combined DFT+U and XL-BOMD framework for shadow QMD simulations was demonstrated with an implementation using the SCC-DFTB LATTE software package for liquid nitromethane and solid nuclear fuel. The combined DFT+U and XL-BOMD approach should be applicable also to a broad range of other methods. The theory in this paper may also demonstrate how similar formulations can be made for other electronic structure methods going beyond regular KS-DFT. Of particular interest are self-interaction corrections 
    \cite{SIC,ULundin01,MPedersen17,JPerdew15}.
    
    \section{Acknowledgements}
    
    This work is supported by the U.S. Department of Energy Office of Basic Energy Sciences (FWP LANLE8AN)
    and by the U.S. Department of Energy through the
    Los Alamos National Laboratory. This research was
    also supported by the Exascale Computing Project (17-SC-20-SC), a collaborative effort of the U.S. Department of Energy Office of Science and the National Nuclear Security Administration.
    Los Alamos National Laboratory is operated by Triad National Security, LLC, for the National Nuclear Security
    Administration of the U.S. Department of Energy Contract No. 892333218NCA000001.
    Discussions with Heather Kulik, Benjamin Hourahine and Joshua Finkelstein are gratefully acknowledged.
        
    \bibliography{mndo_new_xy}
\end{document}